\documentclass[draftclsnofoot,onecolumn,11pt]{IEEEtran}
\usepackage[T1]{fontenc}
\usepackage[latin9]{inputenc}
\usepackage{float}
\usepackage{amsmath}
\usepackage{amssymb}
\usepackage{graphicx}
\usepackage{color}
\makeatletter

\floatstyle{ruled}
\newfloat{algorithm}{tbp}{loa}
\providecommand{\algorithmname}{Algorithm}
\floatname{algorithm}{\protect\algorithmname}

\usepackage{amsfonts}
\usepackage{cite}
\usepackage{array}
\usepackage{algorithm}
\usepackage{algorithmic}
\usepackage{graphicx}

\DeclareMathOperator*{\maximize}{maximize}

\DeclareMathOperator*{\st}{subject~to}

\newcommand{\trans}{^{\mbox{\scriptsize T}}}
\renewcommand{\Im}{\mathrm{Im}}
\makeatother

\begin{document}

\title{Fast Converging Algorithm for Weighted Sum Rate Maximization in Multicell
MISO Downlink}

\author{Le-Nam Tran,~\IEEEmembership{Member,~IEEE}, Muhammad Fainan Hanif, Antti Tölli,~\IEEEmembership{Member,~IEEE}, and Markku Juntti,~\IEEEmembership{Senior Member,~IEEE}
\thanks{Copyright (c) 2012 IEEE. Personal use of this material is permitted. However, permission to use this material for any other purposes must be obtained from the IEEE by sending a request to pubs-permissions@ieee.org.}
\thanks{This research was supported by Tekes (the Finnish Funding Agency for Technology and Innovation), Nokia Siemens Networks, Renesas Mobile Europe, Elektrobit, Xilinx, and Academy of Finland.
}
\thanks{The authors are with the Dept.\ Communications Eng.
and Centre for Wireless Communications, University of Oulu, Finland.
Email: \{ltran, mhanif, atolli, markku.juntti\}@ee.oulu.fi.
}}
\maketitle
\begin{abstract}
The problem of maximizing weighted sum rates in the downlink of a
multicell environment is of considerable interest. Unfortunately,
this problem is known to be NP-hard. For the case of multi-antenna
base stations  and single antenna mobile terminals, we devise
a low complexity, fast and provably convergent algorithm that locally
optimizes the weighted sum rate in the downlink of the system. In
particular, we derive an iterative second-order cone program formulation
of the weighted sum rate maximization problem. The algorithm converges
to a local optimum within a few iterations. Superior performance of
the proposed approach is established by numerically comparing it to
other known solutions.
\end{abstract}

\section{Introduction}
For multiple-input multiple-output (MIMO) broadcast channels, dirty
paper coding (DPC) is known to be the capacity-achieving scheme \cite{Weingarten:CapacityRegion:MU_MIMO:2006}.
However, DPC is a nonlinear interference cancellation technique and
thus requires high complexity. Hence, linear precoding techniques
are of practical interest. Herein, we consider the problem of weighted sum rate maximization
(WSRM) with linear transmit precoding for multicell multiple-input
single-output (MISO) downlink. Unfortunately, the WSRM problem, even
for single-antenna receivers as considered
in this letter, has been shown to be NP-hard in \cite{Luo:SpectrumManagement:2008}.
Although optimal beamformers can be obtained using the methods
presented, for instance, in
\cite{Joshi:MISO:Downlink:BB:2012,Emil:RobustBB:Multicell:2012,Liu:Optimal:MISO:Interference:2012},
they may not be practically useful since the complexity of finding
optimal designs grows exponentially with the problem size. Hence,
the need of computationally conducive suboptimal solutions to the
WSRM problem still remains.

Since the WSRM problem is nonconvex and NP-hard, there exists a class of beamformer designs which are based on  achieving the \emph{necessary}
optimal conditions of the WSRM problem. In fact, this philosophy has been used, e.g., in
\cite{Venturino:CoordinatedBF:2010,Chris:LinearPrecoding:2010,Christensen:WSRM:2008,Shi:WMMSE:2011}.
Interestingly, in \cite{Joshi:MISO:Downlink:BB:2012}, the authors
have numerically shown that the suboptimal designs that achieve the
necessary optimal conditions of the WSRM problem perform very close
to the optimal design. In
\cite{Venturino:CoordinatedBF:2010}, the iterative coordinated
beamforming algorithm was proposed by manipulating the
Karush\textendash{}Kuhn\textendash{}Tucker (KKT) equations. However,
this algorithm is not provably convergent. In
\cite{Christensen:WSRM:2008,Shi:WMMSE:2011}, the WSRM problem with joint transceiver design is solved  using alternating
optimization between transmit and receive beamforming.
As we show by numerical results, these methods have a slower
convergence rate compared to our proposed design.

In this letter, we propose a fast converging algorithm that \emph{locally}
solves the problem of WSRM for multicell MISO downlink. The idea of our
iterative beamformer design is based on the framework of successive convex
approximation (SCA) presented in \cite{Beck:SCA:2010}. The numerical
results show that the proposed algorithm converges within a few
iterations to a locally optimal point of the WSRM problem. The general concept of the SCA method is as follows. In each step of an iterative procedure,
we approximate the original nonconvex problem by an efficiently
solvable convex program and then update the variables involved
until convergence. We note that in the context of transmit linear precoding for
multicell downlink, the SCA method has been used, for example, in
\cite{Chris:LinearPrecoding:2010}. Basically, this method is based on convex relaxations of
the rate function and generally arrives at more complex formulations.
By proper transformations, we approximate the WSRM problem
as a second-order cone program (SOCP) in each step of the SCA method. Our numerical results show that the proposed algorithm generally performs better than the known approaches, in particular, in terms of convergence rate.

\emph{Notation}: We use standard notations in this letter. Bold
lower and upper case letters represent vectors and matrices,
respectively; $\ensuremath{(.)\trans}$ represents the transpose
operator.  $\mathbb{C}^{a\times b}$ represents the space of complex
matrices of dimensions given as superscripts; $|c|$ represents the
absolute value of a complex number. Finally, $\|.\|_{2}$ represents
the $l_{2}$ norm.
\section{\label{PF} Problem Formulation}
Consider a system of $B$ coordinated BSs of $N$ transmit antennas
each and $K$ single-antenna receivers. The set of all $K$ users
is denoted by $\mathcal{U}=\{1,\,2\,\ldots,K\}$. We assume that data
for the $k$th user is transmitted only from one BS, which is denoted
by $b_{k}\in\mathcal{B}$, where $\mathcal{B}\triangleq\{1,2,\ldots,B\}$
is the set of all  BSs. The set of all users served by BS $b$
is denoted by $\mathcal{U}_{b}$. Under flat fading channel conditions,
the signal received by the $k$th user is
 \begin{equation}
 y_{k}  = \mathbf{h}_{b_{k},k}\mathbf{w}_{k}d_{k}+\sum\limits _{i=1,i\neq k}^{K}\mathbf{h}_{b_{i},k}\mathbf{w}_{i}d_{i}+n_{k}\label{eq:system_model}
\end{equation}
where $\mathbf{h}_{b_i,k}\in\mathbb{C}^{1\times N}$ is the channel
(row) vector from BS $b_i$ to user $k$, $\mathbf{w}_{k}\in\mathbb{C}^{N\times1}$
is the beamforming vector (beamformer) from BS $b_{k}$ to user $k$,
$d_{k}$ is the normalized complex data symbol, and $n_{k}\sim\mathcal{C}\mathcal{N}(0,\sigma^{2})$
is complex circularly symmetric zero mean Gaussian noise with variance $\sigma^{2}$. The term
$\sum_{i=1,i\neq k}^{K}\mathbf{h}_{b_{i},k}\mathbf{w}_{i}d_{i}$ in
\eqref{eq:system_model} includes both intra- and inter-cell interference.
The total power transmitted by BS $b$ is $\sum_{k\in\mathcal{U}_{b}}\bigl\|\mathbf{w}_{k}\bigr\|_{2}^{2}$.
The SINR $\gamma_{k}$ of user $k$ is
\begin{equation}
\gamma_{k}=\frac{\bigl|\mathbf{h}_{b_{k},k}\mathbf{w}_{k}\bigr|^{2}}{\sigma^{2}+\sum_{i=1,i\neq k}^{K}\bigl|\mathbf{h}_{b_{i},k}\mathbf{w}_{i}\bigr|^{2}}.
\end{equation}
In this letter, we are interested in the problem of WSRM under per-BS
power constraints%
\footnote{It is straightforward to extend the proposed algorithm to handle per-antenna
power constraints at each BS.%
}, which is formulated as
\begin{equation}
\begin{array}{rl}
\underset{\mathbf{w}_{k}}{\maximize} & \sum_{k=1}^{^{K}}\alpha_{k}\log_{2}(1+\gamma_{k})\\
\st & \sum_{k\in\mathcal{U}_{b}}\|\mathbf{w}_{k}\|_{2}^{2}\leq P_{b},\;\forall b\in\mathcal{B}
\end{array}\label{prob}
\end{equation}
where $\alpha_{k}$'s are positive weighting factors which are typically
introduced to maintain a certain degree of fairness among users. As
mentioned earlier, since problem \eqref{prob} is NP-hard, the globally
optimal design mainly plays as a theoretical benchmark rather than
a practical solution \cite{Emil:RobustBB:Multicell:2012}. Herein, we propose a low-complexity algorithm that solves \eqref{prob}
locally, i.e, satisfies the \emph{necessary} optimal conditions of
\eqref{prob}.
\section{\label{PA} Proposed Low-complexity Beamformer Design}
To arrive at a tractable solution, we note that following
monotonicity of logarithmic function, \eqref{prob} is equivalent to
\begin{equation}
\begin{array}{rl}
\underset{\mathbf{w}_{k}}{\maximize} & \prod_{k}(1+\gamma_{k})^{\alpha_{k}}\\
\st & \sum_{k\in\mathcal{U}_{b}}\|\mathbf{w}_{k}\|_{2}^{2}\leq P_{b},\;\forall b\in\mathcal{B}
\end{array}\label{prob_eq1}
\end{equation}
which can be equivalently recast as \begin{IEEEeqnarray}[\renewcommand{\IEEEeqnarraymathstyle}{\textstyle}]{rl}\label{prob_eq2}
\underset{{\mathbf{w}}_{k},t_{k}}{\maximize}&\quad\prod_{k}t_{k}\IEEEyessubnumber\label{prob_eq20}\\
\st &\quad\gamma_{k}\geq t^{1/\alpha_k}_{k}-1,\; \forall k\in \mathcal{U}\IEEEyessubnumber\label{prob_eq21}\\
&\quad\sum_{k\in\mathcal{U}_{b}}\|{\mathbf{w}}_{k}\|_{2}^{2}\leq P_{b},\;\forall b\in\mathcal{B}.\IEEEyessubnumber\label{prob_eq22}
\end{IEEEeqnarray} The equivalence of \eqref{prob_eq1} and \eqref{prob_eq2} 
can be easily recognized by noting the fact that all constraints in \eqref{prob_eq21} are active at the optimum. Otherwise, we can obtain a strictly larger objective by increasing $t_{k}$ without violating the constraints.
Next, by introducing additional slack variables $\beta_{k}$, we can
reformulate \eqref{prob_eq2} as
\begin{IEEEeqnarray}[\renewcommand{\IEEEeqnarraymathstyle}{\textstyle}]{rl}\label{prob_eq3}\textstyle
\underset{{\mathbf{w}}_{k},t_{k},\beta_{k}}{\maximize}&\quad\prod_{k}t_{k},\IEEEyessubnumber\label{prob_eq30}\\
\st &\quad{\mathbf{h}_{b_{k},k}{\mathbf{w}}_{k}}\geq\sqrt{t^{1/\alpha_k}_{k}-1}\beta_{k},\forall k\in \mathcal{U},\IEEEyessubnumber\label{prob_eq31}\\
&\quad\Im{({\mathbf{h}}_{b_{k},k}{\mathbf{w}}_{k})}=0,\forall k\in \mathcal{U},\IEEEyessubnumber\label{prob_eq311}\\
&\quad \Bigl({ \sigma^{2}+\sum_{i\neq  k}|{\mathbf{h}}_{b_{i},k}{\mathbf{w}}_{i}|^{2}\Bigr)^{1/2}\leq\beta_{k},\forall k\in \mathcal{U},\IEEEeqnarraynumspace\IEEEyessubnumber\label{prob_eq32}}\\
&\quad\sum_{k\in\mathcal{U}_{b}}\|{\mathbf{w}}_{k}\|_{2}^{2}\leq P_{b},\;\forall b\in \mathcal{B}.\IEEEyessubnumber\label{prob_eq33}
\end{IEEEeqnarray}

The equivalence between \eqref{prob_eq2} and \eqref{prob_eq3} is
justified as follows. First, we note that forcing the imaginary part
of $\mathbf{h}_{b_{k},k}\mathbf{w}_{k}$ to zero in
\eqref{prob_eq311} does not affect the optimality of
\eqref{prob_eq2} since a phase rotation on $\mathbf{w}_{k}$ will
result in the same objective while
satisfying all constraints.
 Second, we can show that all the constraints in \eqref{prob_eq32}
hold with equality at the optimum. Suppose, to the contrary, the
constraint for some $k$ in \eqref{prob_eq32} is inactive. Let us
define $\tilde{\beta}_{k}\triangleq\beta_{k}/\eta$ and
$\tilde{t}_{k}\triangleq\{\eta^{2}(t_{k}^{1/\alpha_{k}}-1)+1\}^{\alpha_{k}}$,
where $\eta$ is a positive scaling factor. Since the constraint \eqref{prob_eq32} is inactive, we
can choose $\eta>1$ such that the constraints in \eqref{prob_eq31}
and \eqref{prob_eq32} are still met if we replace
$(\beta_{k},t_{k})$ by $(\tilde{\beta}_{k},\tilde{t}_{k})$. However,
such a substitution results in a strictly larger objective because
$\tilde{t}_{k}>t_{k}$ for $\eta>1$. This contradicts the fact that
we have obtained an optimal solution.

As a step toward a low-complexity solution to the WSRM problem, we
rewrite the constraint \eqref{prob_eq31} as
\begin{IEEEeqnarray}{rCl}\label{prob_eq3new}
\mathbf{h}_{b_{k},k}\mathbf{w}_{k}&\geq&\sqrt{x_k}\beta_{k},\;\forall k \in \mathcal{U}
\IEEEyessubnumber\label{prob_eq3new1}\\ x_k+1 &\geq &t^{1/\alpha_k}_{k},\;\forall k \in \mathcal{U}.
\IEEEyessubnumber\label{prob_eq3new2}\end{IEEEeqnarray}
 Again, we can easily see that by replacing \eqref{prob_eq31} with
\eqref{prob_eq3new1} and \eqref{prob_eq3new2}, we obtain an
equivalent formulation of \eqref{prob_eq3}. The reason of doing so becomes clear shortly. Let us define
$f(x_{k},\beta_{k})=\sqrt{x_{k}}\beta_{k}$ for $x_{k},\beta_{k}\geq0$ and
 focus on the constraint \eqref{prob_eq3new1}
first. Note that $f(x_{k},\beta_{k})$ is nonconvex on the defined
domain, and thus \eqref{prob_eq3new1} is not a convex
constraint. To deal with nonconvex constraints, we invoke a result of
\cite{Beck:SCA:2010} which shows that if we replace
$f(x_{k},\beta_{k})$ by its \emph{convex upper bound} and
iteratively solve the resulting problem by judiciously updating the
variables until convergence, we can obtain a KKT point of
\eqref{prob_eq3}. To this end, for a given $\phi_{k}$ for all $k$,
we define the function \cite{Beck:SCA:2010}
\begin{equation}
G(x_{k},\beta_{k},\phi_{k})\triangleq\tfrac{\phi_{k}}{2}\beta_{k}^{2}+\tfrac{1}{2\phi_{k}}x_{k}\label{def1}
\end{equation}
which arises from the inequality of arithmetic and geometric means
of $\phi_{k}\beta_{k}^{2}$ and
$\phi_{k}^{-1}{x_{k}}$. It is easy to check that
$G(x_{k},\beta_{k},\phi_{k})$ is a convex overestimate of
$f(x_{k},\beta_{k})$ for a fixed $\phi_{k}>0$, i.e.,
$G(x_{k},\beta_{k},\phi_{k})\geq f(x_{k},\beta_{k})$ for all
$\phi_{k}>0$. Moreover, when
$\phi_{k}=\tfrac{\sqrt{x_{k}}}{\beta_{k}}$, it is plain to
observe
\begin{IEEEeqnarray}{rCl}\label{condition}f(x_{k},\beta_{k})
& = & G(x_{k},\beta_{k},\phi_{k})\IEEEyessubnumber\label{defF}\\
\nabla f(x_{k},\beta_{k}) & = & \nabla
G(x_{k},\beta_{k},\phi_{k})\IEEEyessubnumber\label{defG}\end{IEEEeqnarray}
where
$\nabla f$ is the gradient of $f$. Obviously if $f(x_{k},\beta_{k})$
is replaced by $G(x_{k},\beta_{k},\phi_{k})$, \eqref{prob_eq3new1}
can be formulated as a second-order cone (SOC) constraint as we
shown in \eqref{eq:Beck:SOC}.

Now we turn our attention to \eqref{prob_eq3new2}.
 Recall that
$t_{k}^{1/\alpha_{k}}$ is convex if $0<\alpha_{k}\leq 1$ and concave if
$\alpha_{k}>1$, and that the optimal solution of \eqref{prob} stays
the same if we multiply all $\alpha_{k}$'s by the same positive
constant. Thus, we can force \eqref{prob_eq3new2} to be convex by
scaling down $\alpha_{k}$'s in \eqref{prob} such that
$0<\alpha_{k}\leq 1$ for all $k$. However, in this case, the constraint
$x_{k}+1\geq t_{k}^{1/\alpha_{k}}$ cannot
be directly written as an SOC constraint for  $\alpha_{k}\in \mathbb{R}_{++}$.%
\footnote{When $\alpha_{k}$ is an integer or a rational number, we
can transform the constraint \eqref{prob_eq3new2} into a number of
SOC constraints
\cite{Alizadeh:SOCP:2001}. %
} As our goal is to arrive at an SOCP, we instead scale $\alpha_{k}$'s
in \eqref{prob} such that $\alpha_{k}>1$ for all $k$ and thus
$t_{k}^{1/\alpha_{k}}$ becomes concave. Again, in the light of \cite{Beck:SCA:2010},
we replace the right side of the inequality in \eqref{prob_eq3new2}
by its upper bound, which now can be obtained by the first order approximation
due to the concavity of $t_{k}^{1/\alpha_{k}}$. Precisely, we have
\begin{equation}
t_{k}^{\tfrac{1}{\alpha_{k}}}\leq{t_{k}^{(n)}}^{\tfrac{1}{\alpha_{k}}}+\tfrac{1}
{\alpha_{k}}{t_{k}^{(n)}}^{\tfrac{1}{\alpha_{k}}-1}(t_{k}-t_{k}^{(n)})\label{approxC}
\end{equation}
where $t_{k}^{(n)}$ denotes the value of variable $t_{k}$ in the
$n$th iteration (i.e., the iteration corresponding to Algorithm \ref{algo:1}
described later). In fact, we have linearized $t_{k}^{1/\alpha_{k}}$
around the operating point $t_{k}^{(n)}$.
With \eqref{approxC}, \eqref{prob_eq3new2} now becomes a linear
inequality. We note that the linear approximation in \eqref{approxC}
is trivially shown to satisfy the conditions in \eqref{defF} and
\eqref{defG} at $t_{k}^{(n)}$. A question naturally arises is
whether the linear approximation in \eqref{approxC} affects the
optimal sum rate. Interestingly, our numerical experiments show that
the WSR obtained with the successive approximation with
$\alpha_{k}>1$ is identical to that when \eqref{prob_eq3new2} is
forced to be convex by having $0<\alpha_{k}<1$ in (3) for all $k$.
In terms of complexity, the linear inequality in \eqref{approxC} is
more preferable since it requires lower computational effort
compared to the original nonlinear equality constraint in
\eqref{prob_eq3new2}.

Replacing the right sides of \eqref{prob_eq3new1} and \eqref{prob_eq3new2}
by the upper bounds in \eqref{def1} and \eqref{approxC},
respectively, we can formulate \eqref{prob_eq3} as an SOCP by noting
that the objective in \eqref{prob_eq3}, i.e., the product of
$t_{k}$'s admits an SOC representation
\cite{Lobo:SOCP:1998,Alizadeh:SOCP:2001}. The main ingredient in
arriving at the SOCP representation is the fact that the hyperbolic
constraint $uv\geq z^{2}$ where $u\geq0$, $v\geq0$ is equivalent to
$\|[2z\;\;(u-v)]\trans\|_{2}\leq(u+v)$. Let us illustrate the SOCP
formulation of \eqref{prob_eq3} for the special case $K=2^{q}$,
where $q$ is some positive integer. By collecting two variables at a
time and incorporating the additional hyperbolic constraint
corresponding to them, we rewrite \eqref{prob_eq3} as the SOCP in
\eqref{eq:SOCP}, shown on the top of the  page,
\begin{figure*}
\setlength{\IEEEnormaljot}{0.25\normalbaselineskip}\begin{IEEEeqnarray}[\renewcommand{\IEEEeqnarraymathstyle}{\textstyle}]{rl}\label{eq:SOCP}\underset{\mathbf{w}_{k},t_{k},x_k,\beta_{k},z_{k}^{i}}{\maximize}   &\quad z^{(0)}\IEEEyessubnumber\\ \st   &\quad \bigl\|\bigl[2z_{i}^{(N-1)}\quad(t_{2i-1}-t_{2i})\bigr]\trans\bigr\|_{2}\leq(t_{2i-1}+t_{2i}),\quad i=1,\ldots,2^{N-1}\IEEEyessubnumber\\ 
&\quad \qquad\qquad\cdots\cdots\nonumber\\    &\quad \bigl\|\bigl[2z^{(0)}\quad(z_{1}^{(1)}-z_{2}^{(1)})\bigr]\trans\bigr\|_{2}\leq(z_{1}^{(1)}+z_{2}^{(1)}),\IEEEyessubnumber\\    &\quad \bigl\Vert\bigl[\tfrac{1}{2}\bigl(\mathbf{h}_{b_{k},k}\mathbf{w}_{k}-\tfrac{1}{2\phi_{k}^{(n)}}x_{k}-1\bigr)\quad\sqrt{\tfrac{\phi_{k}^{(n)}}{2}}\beta_{k}\bigr]\trans\bigr\Vert_2\leq\tfrac{1}{2}\bigl(\mathbf{h}_{b_{k},k}\mathbf{w}_{k}-\tfrac{1}{2\phi_{k}^{(n)}}x_{k}+1 \bigr),\forall k\in\mathcal{U}\IEEEyessubnumber\IEEEeqnarraynumspace\label{eq:Beck:SOC}\\
&\quad{t_{k}^{(n)}}^{1/\alpha_{k}}+\frac{1}{\alpha_{k}}{t_{k}^{(n)}}^{1/\alpha_{k}-1}(t_{k}-t_{k}^{(n)}) \leq x_{k}+1,\;\forall k\in\mathcal{U},\IEEEeqnarraynumspace\IEEEyessubnumber\label{eq:exp:SOC}\\    
&\quad\bigl\Vert\bigl[\sigma\quad \mathbf{h}_{b_{1},k}\mathbf{w}_{1}\,\cdots\,\mathbf{h}_{b_{k-1},k}\mathbf{w}_{k-1}\quad\mathbf{h}_{b_{k+1},k}\mathbf{w}_{k+1}\,\cdots\,\mathbf{h}_{b_{K},k}\mathbf{w}_{K}\bigr]\trans\bigr\Vert_2\leq\beta_{k},\;\forall k\in\mathcal{U},\IEEEyessubnumber\\
&\quad\sum_{k\in\mathcal{U}_{b}}\|\mathbf{w}_{k}\|_{2}^{2}\leq
P_{b},\;\forall
b\in\mathcal{B}.\IEEEyessubnumber\label{power:constraint}
\end{IEEEeqnarray}\hrulefill
\end{figure*}
where $\phi_{k}^{(n)}$ is the value of $\phi_{k}$ in the $n$th
iteration. 
In the case of $K\neq2^{q}$, we define additional $t_{j}=1$ for
$j=K+1,\ldots,2^{\left\lceil \log_{2}K\right\rceil }$, where
$\left\lceil x\right\rceil $ is the smallest integer not less than
$x$ and the above expression still holds \cite{Alizadeh:SOCP:2001}.
Now we are in a position to present an algorithm that solves problem
\eqref{prob} locally. The pseudocode of the beamformer design is
outlined in Algorithm~\ref{algo:1}.
\begin{algorithm}[b]
{\small \begin{algorithmic}[1] \caption{ {\small Proposed beamformer design for the WSRM problem in multicell MISO
downlink.}}
\label{algo:1}
\renewcommand{\algorithmicrequire}{\textbf{Initialization:}}
\REQUIRE $n=0$, $\ensuremath{(\phi_{k}^{(n)},t_{k}^{(n)})=\textrm{random}}$.
\REPEAT
\STATE Solve \eqref{eq:SOCP} with $\phi_{k}^{(n)}$ and $t_{k}^{(n)}$, and denote optimal
$(t_{k},\beta_{k},x_{k})$ as $(t_{k}^{\star},\beta_{k}^{\star},x_{k}^{\star})$.
\STATE Update $(t_{k}^{(n+1)},\beta_{k}^{(n+1)},x_{k}^{(n+1)})=(t_{k}^{\star},\beta_{k}^{\star},x_{k}^{\star})$
and $\phi_{k}^{(n+1)}=\frac{\sqrt{x_{k}^{(n+1)}}}{\beta_{k}^{(n+1)}}$;
$n:=n+1$. \label{algo:update}
\UNTIL convergence
\end{algorithmic}}
\end{algorithm}
We now present the convergence analysis of Algorithm~\ref{algo:1}.
Consider the $n+1$st iteration of Algorithm~\ref{algo:1} that solves the
optimization problem \eqref{eq:SOCP}. If we replace
$(t_{k},\beta_{k},x_{k})$ by
$(t_{k}^{(n)},\beta_{k}^{(n)},x_{k}^{(n)})$ and $\mathbf{w}_{b,k}$
by $\mathbf{w}_{b,k}^{(n)}$, all the constraints in
\eqref{eq:Beck:SOC}-\eqref{power:constraint} are still satisfied.
That is to say, the optimal solution of the $n$th iteration is a
feasible point of the problem in the $n+1$st iteration. Thus, the
objective obtained in the $n+1$st iteration is larger than or equal
to that in the $n$th iteration. In other words,
Algorithm~\ref{algo:1} generates a nondecreasing sequence of
objective values. Moreover, the problem is bounded above due to the
power constraints. Hence, Algorithm \ref{algo:1} converges to some
local optimum solution of \eqref{eq:SOCP}. By the two properties
shown in \eqref{condition} and based on the arguments presented in
\cite{Beck:SCA:2010}, it can be shown that this solution also
satisfies the KKT conditions of \eqref{prob_eq3}. Numerical results
in Section \ref{NR} confirm that Algorithm~\ref{algo:1} performs
very close to {\color{black}optimal linear  design}.

\section{\label{NR}Numerical Results}
\begin{figure}
\centering\includegraphics[width=1\columnwidth]{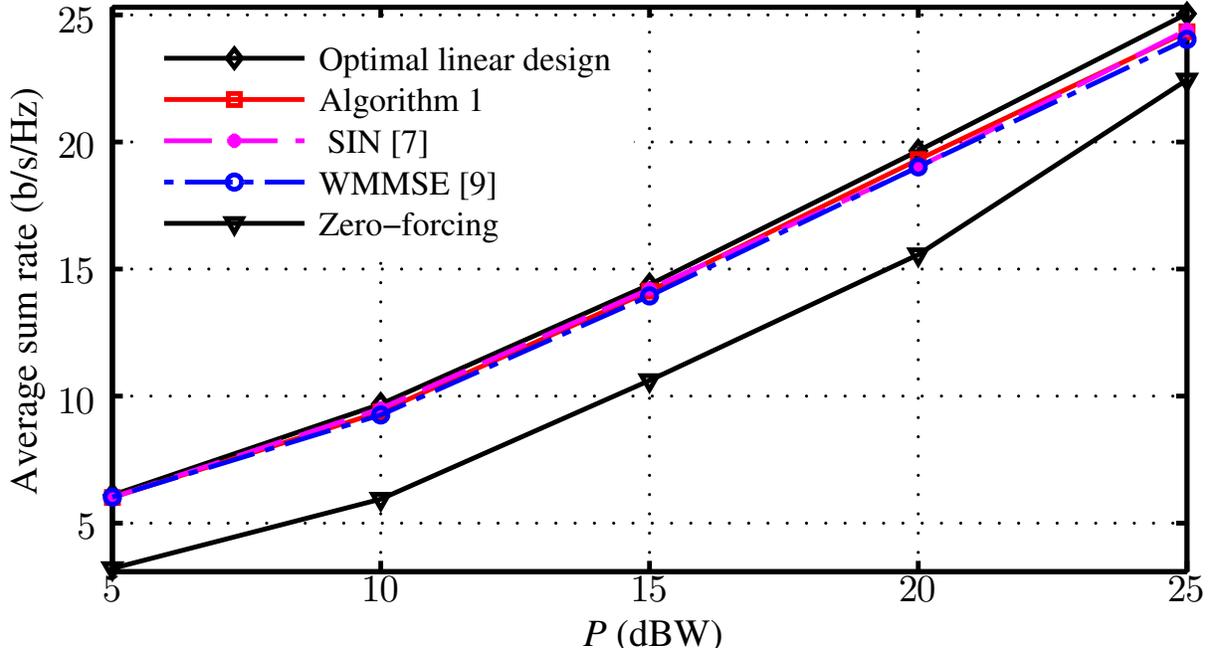}\caption{Average
sum rate comparison for single-cell scenario, $N=4$, $K=4$.}
\label{fig:AverageSR}
\end{figure}
In this section, we numerically evaluate the performance of
Algorithm \ref{algo:1} under different setups using YALMIP
\cite{YALMIP} with SDPT3 \cite{sdpt3} as internal solver. In the
first experiment, we consider a single-cell scenario where a BS with
$N=4$ transmit antennas serves $K=4$ users. The entries of
$\mathbf{h}_{b,k}$ are $\mathcal{CN}(0,1)$ and the noise variance $\sigma^2=1$. In
Fig.~\ref{fig:AverageSR}, we plot the average sum rate
($\alpha_{k}=1$ for all $k$) versus the total transmit power $P$ at
the BS. The achieved sum rate of Algorithm \ref{algo:1} is compared
to those of  zero-forcing
beamforming \cite{Spencer:BD:2004}, the weighted sum mean-square
error minimization (WMMSE) algorithm in \cite{Shi:WMMSE:2011},
the soft inference nulling (SIN) scheme in
\cite{Chris:LinearPrecoding:2010}, {\color{black} and the optimal linear design using the branch-and-bound (BB) method in \cite{Emil:RobustBB:Multicell:2012,Joshi:MISO:Downlink:BB:2012}}.  Initial values for beamformers for the suboptimal schemes
in \cite{Shi:WMMSE:2011,Chris:LinearPrecoding:2010}, and
$(\phi_{k}^{(0)},t_{k}^{(0)})$ in Algorithm \ref{algo:1} are generated randomly.
The sum rate is obtained after Algorithm \ref{algo:1} and the
iterative suboptimal schemes in \cite{Shi:WMMSE:2011,Chris:LinearPrecoding:2010} converge, i.e., the increase in
the objective value between two consecutive iterations is less then
$10^{-2}$. {\color{black} The gap tolerance between the upper and lower bounds for the BB method  is set to $10^{-1}$ as in \cite{Emil:RobustBB:Multicell:2012,Joshi:MISO:Downlink:BB:2012}, and the resulting sum rate  is calculated as the average of the upper and lower bounds}.\footnote{{\color{black} The principle of the BB method to compute an optimal solution to  nonconvex problems is to find provable lower and upper bounds on the globally optimal value and guarantee that the bounds converge as iterations evolve.}}  Results reveal that the average sum rate of Algorithm \ref{algo:1} and other iterative beamformer designs is  the same on convergence and close
to that of {\color{black}the optimal linear approach}. However, the  SIN scheme, WMMSE algorithm and {\color{black}the optimal design} have a slower convergence rate as discussed next.

\begin{figure}
\centering\includegraphics[width=1\columnwidth]{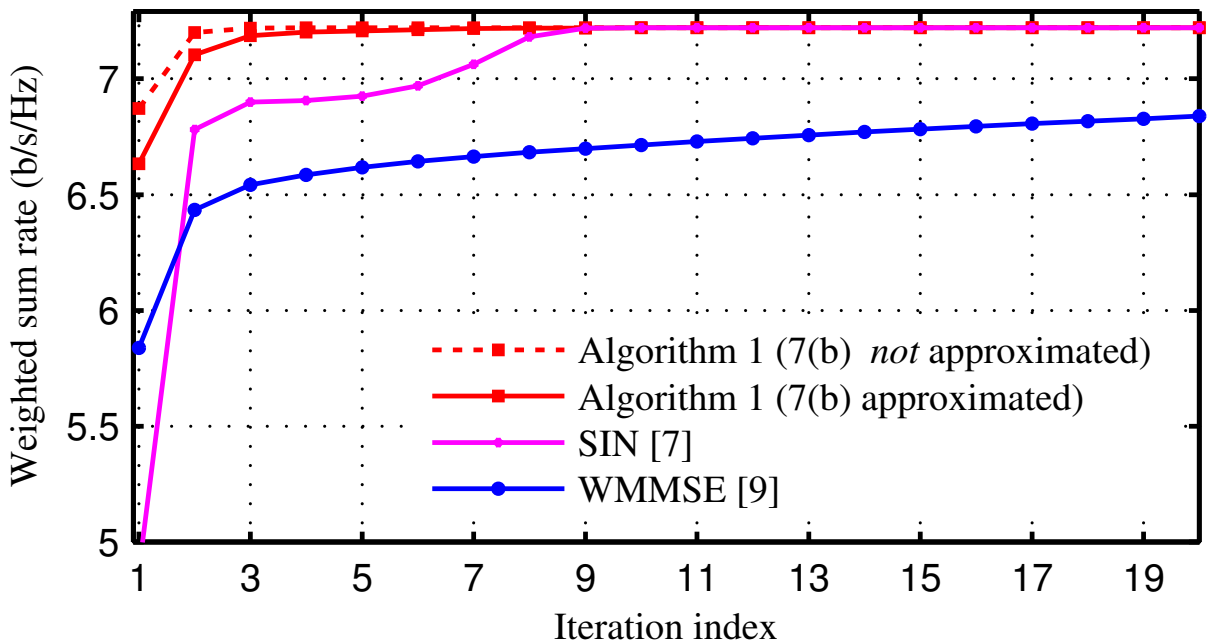}\caption{Convergence
rate of the weighted sum rate, $B=2$, $N=8$, $K=4$, $P_b=12$ dB for all $ b$.}
\label{fig:WSR:convergence}
\end{figure}

In the second experiment, we illustrate the convergence rate of all
considered iterative suboptimal schemes. A simple two-cell scenario
with each BS serving $2$ users is considered. The number of transmit
antennas at each BS is set to $N=8$. The weights, without loss of generality, are taken as
$(\alpha_{1},\alpha_{2},\alpha_{3},\alpha_{4})=(0.14,0.21,0.28,0.36)$
and the power budget of each BS is set to $P_{b}=12$ dB for $b=1,2$.
Fig. \ref{fig:WSR:convergence} compares the weighted sum rate of the
considered schemes as a function of iterations needed to obtain a
stabilized output for a random channel realization. In particular,
our algorithm has converged just after a few of iterations, while
the WMMSE algorithm is still less than midway to convergence. In
fact, for this particular case the WMMSE took hundreds of runs
before converging to the local optimum solution. This observation
may be attributed to the fact that optimization strategy of
\cite{Shi:WMMSE:2011} requires alternate updates between transmit
and receive beamformers and therefore exhibits slower convergence
properties. We  have also noticed that for certain initial values the convergence rate of the WMMSE algorithm is greatly improved. In fact, it was reported in \cite{Bezdek:CAO:2003} that  the convergence rate of an alternating optimization algorithm depends on the initial values of the  variables involved, and it  converges quickly if  initial guess is relatively close to the optimal solution. Further, we observe that Algorithm \ref{algo:1} with and
without scaling has slightly different convergence rate (labeled in
Fig. \ref{fig:WSR:convergence} as `approximated' and `not approximated') but same optimal
value. This validates that the approximation used to arrive at an
SOCP formulation has no impact on the achieved sum rate. Our numerical results reveal
that for other channel realizations, the SIN scheme may have similar
convergence behavior to Algorithm \ref{algo:1}, but the average per
iteration running time of Algorithm \ref{algo:1} is approximately
four times less than that of the SIN method. {\color{black}For the set of channel realizations considered in Fig. \ref{fig:WSR:convergence}, the optimal design also converges to  the same point achieved by other iterative suboptimal methods. However, it takes more than $600$ iterations to reduce the gap between lower and upper bounds of the BB method to less than $10^{-1}$}. Theoretically, the faster
convergence of Algorithm \ref{algo:1} may be attributed to solving
an explicit SOCP in each of its iterations. The faster convergence
of our algorithm can be much useful for distributed implementation
which is left as future work.

\section{\label{Con}Conclusion}
In the letter we have studied the problem of WSRM in the downlink of
multicell MISO system. Since the problem is NP-hard, we have proposed
a low-complexity approximation of the optimization problem. We show
that the problem can be approximated by an iterative SOCP procedure.
While the convergence of the algorithm can be proved, its global optimality
cannot be established. Nonetheless, the algorithm outperforms the
previously studied solutions to the WSRM problem, in particular, in
terms of its convergence rate.


\end{document}